\documentclass[nofootinbib, aps,11pt,preprintnumbers]{revtex4-1}

\usepackage{graphicx}
\usepackage{amssymb}
\usepackage{textcomp}
\usepackage{amsmath}
\usepackage{bm}
\usepackage{times}
\usepackage{epsfig}
\usepackage{color}
\usepackage{array,multirow}
\usepackage{ulem}

\newcommand{\trix}[1]{\left(\begin{array}{#1}}
\newcommand{\notrix}{\end{array}\right)}
\newcommand{\comment}[1]{}

\def\beq{\begin{equation}}
\def\eeq{\end{equation}}
\def\bea{\begin{eqnarray}}
\def\eea{\end{eqnarray}}

\begin{document}
\title{\Large  {{\bf{Supersymmetric Sneutrino-Higgs Inflation}}}}
\author{{Rehan Deen$^{1}$, Burt A.~Ovrut$^{1}$ and Austin Purves$^{1,2}$} \\[2mm]
    {\it $^{1}$ Department of Physics, University of Pennsylvania} \\
   {\it Philadelphia, PA 19104--6396}\\
   {\it $^{2}$ Department of Physics, Manhattanville College}\\
   {\it Purchase, NY 10577} \\[4mm]
}
\date{\today}

{\let\thefootnote\relax\footnotetext{\mbox{ovrut@elcapitan.hep.upenn.edu,  ~Austin.Purves@mville.edu, ~rdeen@sas.upenn.edu~} }}
\setcounter{footnote}{0}

\begin{abstract}
\ \\ [10mm]
{\bf ABSTRACT:} 
It is shown that in the phenomenologically realistic supersymmetric $B-L$ MSSM theory, a linear combination of the neutral, up Higgs field with the third family left-and right-handed sneutrinos can play the role of the cosmological inflaton. Assuming that supersymmetry is softly broken at a mass scale of order $10^{13}~\mathrm{GeV}$, the potential energy associated with this field allows for 60 e-foldings of inflation with the cosmological parameters being consistent with all Planck2015 data. The theory does not require any non-standard coupling to gravity and the physical fields are all sub-Planckian during the inflationary epoch. It will be shown that there is a ``robust'' set of initial conditions which, in addition to satisfying the Planck data, simultaneously are consistent with all present LHC phenomenological requirements.
	
\end{abstract}

\maketitle

\section{Introduction}
In a series of papers \cite{Braun:2005nv, Ovrut:2012wg, Marshall:2014kea, Marshall:2014cwa, Ovrut:2014rba, Ovrut:2015uea, Barger:2008wn, Everett:2009vy, FileviezPerez:2008sx, FileviezPerez:2012mj}, it was shown that a phenomenologically realistic N=1 supersymmetric theory--the $B-L$ MSSM--can emerge as the low energy limit of the observable sector of a class of vacua of heterotic M-theory \cite{Lukas:1998yy, Lukas:1998tt}. Below the GUT compactification scale of order $3 \times 10^{16}$ GeV, this theory has {\it exactly} the spectrum of the MSSM with three right-handed neutrino chiral multiplets, one per family, and gauge group $SU(3)_{C} \times SU(2)_{L} \times U(1)_{Y} \times U(1)_{B-L}$. It is only the gauged  $U(1)_{B-L}$ symmetry that differentiates this theory from the MSSM; hence the name $B-L$ MSSM.  
In the analyses presented in \cite{Braun:2005nv, Ovrut:2012wg, Marshall:2014kea, Marshall:2014cwa, Ovrut:2014rba, Ovrut:2015uea, Barger:2008wn, Everett:2009vy, FileviezPerez:2008sx, FileviezPerez:2012mj}, spacetime was taken to be flat and all superfields had canonically normalized kinetic energy. Without specifying their origin, it was assumed that supersymmetry (SUSY) was broken by the complete set of soft SUSY breaking operators \cite{Girardello:1981wz}. All renormalization group equations (RGEs) were then presented and solved, 
subject to a statistical scan over a selected region of initial SUSY breaking parameter space. For soft SUSY breaking mass parameters of order $1-10$ TeV, it was shown that all low energy phenomenological requirements--that is, breaking $B-L$ and electroweak (EW) symmetry at appropriate values, satisfying all lower bounds on sparticles masses and giving the lightest neutral Higgs mass to be $\sim$125 GeV--were obtained for a {\it robust} set of points in the initial parameter space. An important assumption in the above analysis is that the masses of all geometrical and vector bundle moduli are sufficiently heavy to be integrated out of the effective  theory.

In this paper, we again consider the $B-L$ MSSM theory; not, however,  focussing on its realistic low energy phenomenology but, rather,  as a possible natural framework for a theory of inflation satisfying all recent Planck2015 bounds \cite{Ade:2015lrj}. To do this, we must begin by coupling the theory to $N=1$ supergravity.
Recall that the $B-L$ MSSM arises as the observable sector of  a specific set of vacua of heterotic M-theory \cite{Braun:2005nv}. The coupling of the four-dimensional observable sector of a generic M-theory compactification was carried out in \cite{Lukas:1997fg}.
This result is easily used to determine the explicit Lagrangian for the $B-L$ MSSM coupled to N=1 supergravity. The results are the following.

First, in the limit that the reduced Planck mass $M_{P} \rightarrow \infty$, the resulting theory is precisely the spectrum and Lagrangian of the flat space $B-L$ MSSM--with an important addition. The compactification from eleven to four-dimensions potentially introduces ``moduli'' fields into the low energy theory. These correspond to the geometrical moduli of the Calabi-Yau threefold, the one geometrical radial modulus of the $S^{1}/{{\mathbb{Z}}_{2}}$ orbifold and the moduli of the $SU(4)$ vector bundle \cite{Donagi:1999jp, Buchbinder:2002ji, Buchbinder:2002pr, Braun:2005zv}. It is expected that all moduli develop a non-perturbative potential energy which fixes their vacuum expectation values (VEVs) and gives them mass. In this paper, we will assume that--with the exception of the two complex ``universal'' geometrical moduli-- all of them are sufficiently heavy that they will not appear in the low energy theory. The real parts of the two universal moduli are the ``breathing'' modes of the CY and the orbifold respectively, and are formally defined to be the $a(x)$ and $c(x)$ fields in the eleven-dimensional metric 
\begin{equation}
ds^{2}=g_{\mu\nu}dx^{\mu}dx^{\nu} + e^{2a(x)}\Omega_{AB} dx^{A}dx^{B} + e^{2c(x)} (dx^{11})^{2}  \ .
\label{burt1}
\end{equation}
In the $M_{P} \rightarrow \infty$ limit, these universal moduli, although less massive, ``decouple'' from ordinary matter and can be ignored. However, for finite $M_{P}$ this is no longer the case, and they must be included when we couple the $B-L$ MSSM to supergravity. Specifically, when the $B-L$ MSSM is coupled to $N=1$ supergravity to order $\kappa^{2/3}$ in heterotic M-theory, the K\"ahler potential for the complex scalar fields is modified to become 
\begin{equation}
K= -{\rm ln}(S + \bar{S}) -3~{\rm ln}(T + \bar{T} - \sum_{i}{ \frac{|C_{i}|^{2}}{M_{P}^{2}}}) \ ,
\label{burt2}
\end{equation}
where the sum is over all complex scalar matter fields $C_{i}$ in the $B-L$ MSSM and 
\begin{equation}
S = e^{6a} + i \sqrt{2} \sigma~,~~T=e^{2\hat{c}} + i \sqrt{2} \chi + \frac{1}{2} \sum_{i}{ \frac{|C_{i}|^{2}}{M_{P}^{2}}} 
\label{burt3}
\end{equation}
with $\hat{c}=c + 2a$. The $\sigma$ and $\chi$ fields arise as the duals of specific forms and are required by supersymmetry to extend $a$ and $c$ to the complex fields $S$ and $T$ respectively. The fact that the second term in \eqref{burt2} is a logarithm of a specific type will play a fundamental role in our analysis. Hence, it is important to note that both the K\"ahler potential $K$ in \eqref{burt2}, as well as the specific field definitions given in \eqref{burt3}, are identical to those found by Witten \cite{Witten:1985xb} within the context of the weakly coupled heterotic string. In addition, it was shown by a number of authors \cite{Cecotti:1987dn, Ovrut:1987gk, Lauer:1987kh} that the off-shell structure of the $N=1$ supergravity multiplet arising in heterotic string theory should be that of so-called ``new minimal'' supergravity. In \cite{Ferrara:1988qxa}, it was demonstrated that K\"ahler potentials of the above logarithmic form are consistent with this requirement. The form of (2) has previously arisen in so-called ``no-scale" supergravity models as discussed in \cite{Cremmer:1983bf, Ellis:1983sf, Lahanas:1986uc}. Finally, we find that to order $\kappa^{2/3}$ in heterotic M-theory, the gauge kinetic function in the observable sector is given by
\begin{equation}
f=S \ .
\label{burt4}
\end{equation}
As with the K\"ahler potential, this form of the gauge kinetic function $f$ is identical to that found in the weakly coupled heterotic string and is consistent with coupling to new minimal supergravity. Henceforth, unless otherwise specified, {\it we will work in units in which $M_{P}=1$}.

Inserting the above expressions for $K$ and $f$ , as well as the superpotential $W$ for the $B-L$ MSSM \cite{Ovrut:2015uea}, into the canonical expression for the Lagrangian of $N=1$ matter/gauge fields coupled to supergravity \cite{Lukas:1997fg}, explicitly realizes our goal of coupling the $B-L$ MSSM theory to $N=1$ supergravity. Using this Lagrangian, we begin our analysis of the $B-L$ MSSM as a potential framework for cosmological inflation. We first note that the moduli fields $S$ and $T$, although appearing in the expressions for $K$ and $f$, are assumed to have constant VEVs. One can then show that by appropriate rescaling of all matter fields, as well as all coupling parameters, that is, setting
\begin{equation}
C^{\prime}_{i} = \big(\frac{3}{T+\bar{T}}\big)^{1/2} C_{i}~, \qquad g^{\prime}_{a}=\big(\frac{2}{S+\bar{S}}\big)^{1/2} g_{a}~,~{~\rm for}~a=3,2,3R,BL^{\prime} \ ,
\label{burt5}
\end{equation}
the $S$ and $T$ constants can be completely eliminated from the effective Lagrangian. Henceforth, we will drop the prime on all fields and couplings. It follows that the form of the effective Lagrangian is unaltered, but that the K\"ahler potential and the gauge kinetic function are now given by
\begin{equation}
K= -3~{\rm ln}(1 - \sum_{i}{ \frac{|C_{i}|^{2}}{3}}) \ ,~\qquad f=1 \ .
\label{burt6}
\end{equation}
Recalling that the matter kinetic energy terms in the Lagrangian are 
\begin{equation}
-K_{i\bar{j}}\partial_{\mu}C^{i}\partial^{\mu}C^{\bar{j}} \ ,
\label{burt7}
\end{equation}
it follows from \eqref{burt6} that for {\it small values of the fields $C_{i}$} the kinetic terms do not mix and are all canonically normalized. This is no longer true, however, for field values approaching the Planck scale. 
We continue by analyzing the remaining parts of the Lagrangian. To begin, we find that the pure gravitational action is simply given by
\begin{equation}
-\frac{1}{2} \int_{M_{4}} \sqrt{-g}R \ .
\label{burt8}
\end{equation}
That is, in this analysis
\begin{itemize}
\item {\it The pure gravitational action is canonical. We do not require any ``non-canonical'' coupling of matter to the curvature tensor $R$.}
\end{itemize}
Now consider the potential energy terms for the matter fields in the effective Lagrangian. These break into three types. The supersymmetric F-term and D-term potentials are given by
\begin{equation}
V_{F}= e^{K}\big( K^{i \bar{j}}D_{i}W \overline{D_{\bar{j}}W}-3|W|^{2}\big)~, \qquad  V_{D}=\frac{1}{2} \sum_{a}D_{a}^{2}
\label{burt9}
\end{equation}
respectively, where $W$ is the $B-L$ MSSM superpotential \cite{Ovrut:2015uea}, the $D_{a}$, $a=3,2,3R,BL^{\prime}$ functions are 
\begin{equation}
D^{r}_{a} = -g_a \frac{\partial K}{\partial C_i} [T_{(a)}^r]_i\,^j C_j 
= \frac{g_a }{\left(1 - \tfrac{1}{3}\sum_i |C_i|^2\right)} \mathcal{D}^{r}_{(a)}~,  \qquad  
\mathcal{D}_{(a)}^{r} = -\overline{C}^i [T_{(a)}^r]_i\,^j C_j
\end{equation}
and $T_{(a)}^{r}$, $r=1,\dots,{\rm dim}~G_{a}$  are the the generators of the group $G_{a}$. For the $B-L$ MSSM we find 
\bea
-\mathcal{D}_{(3)}^r =  ({\overline{u}_{R,\rm f}})^m [\Lambda^r]_m\,^n (u_{R,{\rm f}})_n 
&+&  (\overline{d}_{R,{\rm f}})^m [\Lambda^r]_m\,^n (d_{R,{\rm f}})_n  \nonumber  \\
+  (\overline{u}_{L,{\rm f}})^m [\Lambda^r]_m\,^n (u_{L,{\rm f}})_n
&+&  (\overline{d}_{L,{\rm f}})^m [\Lambda^r]_m\,^n (d_{L,{\rm f}})_n \ ,
\label{D3C}
\eea
\begin{equation}
-\mathcal{D}_{(2)}^r = (\overline{H}_{u})^k [\tau^r]_k\,^l (H_{u})_l + (\overline{H}_{d})^k [\tau^r]_k\,^l (H_{d})_l 
+(\overline{Q}_{\rm f})^k [\tau^r]_k\,^l (Q_{\rm f})_l +(\overline{L}_{\rm f})^k [\tau^r]_k\,^l (L_{\rm f})_l  \ ,
\label{D2L}
\end{equation}
\bea
 -\mathcal{D}_{(3R)} &=& \tfrac{1}{2} \left( |H_{u}^{+}|^2  +  |H_{u}^{0}|^2 -  |H_{d}^{0}|^2 - |H_{d}^{- }|^2 \right)
 \nonumber \\
&& -\tfrac{1}{2} |\nu_{R,{\rm f}}|^2+\tfrac{1}{2} |e_{R,{\rm f}}|^2 - \tfrac{1}{2}|u_{R,{\rm f}}|^2  +\tfrac{1}{2}|d_{R,{\rm f}}|^2  \ , 
\label{D3R}
\eea
\bea
-\mathcal{D}_{(BL^{\prime})} &=& - |\nu_{L,{\rm f}}|^2 - |e_{L,{\rm f}}|^2 + \tfrac{1}{3}|u_{L,{\rm f}}|^2 + \tfrac{1}{3}|d_{L,{\rm f}}|^2 \nonumber \\
&&+ |\nu_{R,{\rm  f}}|^2 + |e_{R,{\rm f}}|^2 - \tfrac{1}{3}|u_{R,{\rm f}}|^2 - \tfrac{1}{3}|d_{R,{\rm f}}|^2.
\label{DB-L}
\eea
The subscript ${\rm f}=1,2,3$  labels the families, the matrices  $\Lambda^r$ and $\tau^r$ are the generators of $SU(3)_C$ and $SU(2)_L$ respectively, while $m, n$, are color indices and $k,l$ are $SU(2)$ indices. In addition, there is a soft supersymmetry breaking potential given by
\begin{eqnarray}
V_{soft} & = & (m^{2}_{Q_{{\rm f}}}|{Q}_{{\rm f}}|^{2}+m^{2}_{u_{R,{\rm{f}}}}|{u}_{R,{\rm f}}|^{2}+
                        m^{2}_{d_{R,{{\rm f}}}}|{d}_{R,{\rm f}}|^{2} 
                  + m^{2}_{L_{{\rm f}}}|{L}_{{\rm f}}|^{2}+m^{2}_{\nu_{R,{\rm f}}}|{\nu}_{R,{\rm f}}|^{2}
                        \label{9}  \nonumber \\   
          & & +m^{2}_{e_{R,{\rm f}}}|{e}_{R,{\rm f}}|^{2})+m_{H_{u}}^{2}|H_{u}|^{2} +m_{{\bar{H}}_{d}}^{2}|{H}_{d}|^{2} + (bH_{u}H_{d} + h.c.) +\dots \ , 
 \label{8}  
\end{eqnarray}
where, for simplicity, we have not shown the cubic scalar terms. Suffice it to say that we assume that each of their  dimensionful coefficients is proportional to the associated Yukawa coupling. In all three potentials, the sum over families is implicit.

It is possible to find solutions for which the D-term potential $V_D$ vanishes, the so-called ``D-flat" directions. Such a solution will be central to the construction of our inflationary potential. 
In this paper, to satisfy the D-flatness condition, we restrict ourselves to fields that are not charged under $SU(3)_C$ or under $U(1)_{EM}$. Hence,  we are naturally lead to the field space configuration 
\begin{equation}
H_u^0 = \nu_{R,3} = \nu_{L,3} \ ,
\label{D-flat}
\end{equation}
with all other fields set to zero. We note that only in a model containing right-handed neutrino superfields, such as the $B-L$ MSSM, would such a D-flat direction arise.
 
Our neutral D-flat direction can be enhanced by the inclusion of the $H_d^0$ field. 
Consider the region of the field space defined by the equation 
\begin{equation}
\sqrt{(1-\beta^2)} H_d^0 = \beta  \sqrt{(1 - \beta^2)} H_u^0 = \beta \nu_{3,L} = \beta \nu_{3,R} .
\end{equation}
This equation defines a one-parameter family of D-flat directions.
For $\beta = 0$, this region gives us the field direction given by equation \eqref{D-flat}, with $H_d^0 = 0$.
For $\beta = 1$, we retrieve the D-flat direction given by $H_d^0 = H_u^0$, with $\nu_{3,R} = \nu_{3,L} = 0$. 

By using the D-flat direction $H_d^0 = H_u^0$, we can repeat our above construction of the inflation potential with essentially identical results. An alternative inflation scenario using this direction was carried out in \cite{Ibanez:2014swa, Bielleman:2015lka}.


\section{Inflation}


\subsection{Inflationary potential}

From our preferred D-flat direction (\ref{D-flat}), we construct an inflationary potential as follows. First, define three new fields $\phi_{i}$, $i=1,2,3$ using
\bea
H_u^0 &=&\tfrac{1}{\sqrt{3}}\left(\phi _1- \phi _2- \phi _3 \right), \nonumber \\
\nu _{L,3} &=& \tfrac{1}{\sqrt{3}}\phi _1+\left(\tfrac{1}{2}+\tfrac{1}{2 \sqrt{3}}\right) \phi _2+\left(\tfrac{1}{2 \sqrt{3}}-\tfrac{1}{2}\right) \phi _3, \nonumber \\
\nu _{R,3} &=& \tfrac{1}{\sqrt{3}}\phi _1+\left(\tfrac{1}{2 \sqrt{3}}-\tfrac{1}{2}\right) \phi _2+\left(\tfrac{1}{2}+\tfrac{1}{2 \sqrt{3}}\right) \phi _3.
\label{redef}
\eea
Recall that all other fields have been set to zero. The field $\phi_1$ corresponds to the D-flat field direction while $\phi_2$ and $\phi_3$ are two orthogonal directions.  One may verify this by restricting attention to this three-dimensional subspace and noting that the D-term potential vanishes when $\phi_2 = \phi_3 =0$ for any value of $\phi_1$. For future reference, we note that 
\begin{equation}
\phi_1 = \tfrac{1}{\sqrt{3}} \left( H_u^0 + \nu _{L,3} + \nu _{R,3}\right)
\label{phi_1}
\end{equation}
and the associated quadratic soft mass squared is given by
\begin{equation}
m^{2}=\frac{1}{3}(m^{2}_{H_{u}^{0} }+ m_{\nu_{L,3}}^{2}+m_{\nu_{R,3}}^{2}) \ .
\label{susy2}
\end{equation}
Setting all fields to zero with the exception of $\phi_{1}$, the $V_{D}$ potential vanishes and the Lagrangian becomes
\begin{equation}
\mathcal{L} = -\frac{1}{\left(1 - \tfrac{1}{3}|\phi_1|^2\right)^2}\partial_\mu{\overline \phi_1} \partial^\mu{\phi_1}  - V_F(\phi_1) - V_{soft} (\phi_1)~, 
\label{lagrange1}
\end{equation}
where
\begin{equation}
V_F(\phi_1) = \frac{3 |\phi _1|^2 \left(\mu ^2+ Y_{\nu3}^2|\phi _1|^2 \right)}{\left(3 - |\phi _1|^2\right)^2}~, \qquad
V_{soft}(\phi_1) = m^2 |\phi_1|^2 \ .
\label{lagrange1a}
\end{equation}
Here $Y_{\nu3}$ is the third-family sneutrino Yukawa coupling and $\mu$ is the usual supersymmetric Higgs parameter\footnote{Note that the contribution from the cubic soft SUSY breaking term to $V_{soft}$ is negligible.}.

Since this Lagrangian is symmetric under global $U(1)$ transformations, we choose our inflaton to be the real $\overline{\phi}_1 = \phi_1$ field, the potential for the imaginary part of $\phi_{1}$  simply being flat. That is, the inflaton is a single real-valued field, which (somewhat abusing notation) we continue to denote by  $\phi_{1}$. We want to emphasize that

\begin{itemize}

\item {\it The inflaton is a linear combination of the real parts of $H_{u}^{0}$, $\nu_{L,3}$ and $\nu_{R,3}$ and, hence, is composed of fields already appearing in the $B-L$ MSSM.}

\end{itemize}

\noindent In order to to canonically normalize the kinetic energy term, we make a field  redefinition to a real scalar $\psi$ given by
\begin{equation}
\phi_1 = \sqrt{3} \tanh\left(\frac{\psi}{\sqrt{6}} \right) .
\label{psi}
\end{equation}
In terms of the new field $\psi$, Lagrangian \eqref{lagrange1} now becomes
\begin{equation}
\mathcal{L} = -\frac{1}{2}\partial_\mu \psi \partial^\mu \psi - V_F(\psi)  - V_{soft}(\psi)
\label{lagrange2} 
\end{equation}
where $V_{F}(\psi)$ is obtained from the first term in  \eqref{lagrange1a} using \eqref{psi} and
\begin{equation}
V_{soft}(\psi)= 3 m^2 \tanh^2\left( \frac{\psi}{\sqrt{6}}\right) \ .
\label{susy3}
\end{equation}

\subsection{The primordial parameters}
For an {\it arbitrary} potential function $V(\psi)$, one defines the ``slow-roll'' parameters to be
\begin{equation}
\epsilon = \tfrac{1}{2}\left(\frac{V^{\prime}}{V}\right)^2 ~, \qquad 
\eta = \frac{V^{\prime\prime}}{V} \ .
\label{slow-roll}
\end{equation}
For there to be an interval of slow-roll inflation, these parameters must satisfy the conditions that $\epsilon, |\eta| \ll 1$. Assuming this to be the case for some range of $\psi$, one defines the end of the slow-roll period to be the smallest value of $\psi$ for which $\epsilon=1$. This will be denoted by $\psi_{end}$. To satisfy the CMB data, it is necessary that there be at least $60$ e-foldings of inflation preceding $\psi_{end}$. The value of the field which precedes $\psi_{end}$ by  {\it exactly} $60$ e-folds is found by integrating the function $1/\sqrt{2\epsilon}$, and will be denoted by $\psi_{*}$. The spectral index $n_s$ and the scalar-to-tensor ratio $r$ are then defined to be
\begin{equation}
n_s \simeq  1 + 2\eta_* - 6 \epsilon_*~,
\qquad r \simeq 16 \epsilon_* \ ,
\label{tel1}
\end{equation}
where the label $``_*"$, here and below,  denotes quantities that are evaluated at $\psi_{*}$. In addition,
the Planck2015 normalization of the CMB fluctuation amplitude requires that the energy scale of inflation satisfies
\begin{equation}
V_*^{1/4} = 1.88\big(\frac{r}{0.10}\big)^{1/4}  \times 10^{16}~\mathrm{GeV}  ~,
\label{Vconstraint}
\end{equation}
where we have restored dimensionful units for clarity.

With this in mind, let us analyze our specific potential $V=V_{F}+V_{soft}$ presented above. We begin by considering $V_{soft}$  in \eqref{susy3} alone, 
momentarily ignoring $V_F$. We find that the requirement of 60 e-folds of inflation leads to the results that
\begin{equation}
 \psi_{end} = ~ 1.21~, \qquad\psi_* = 6.23  \ .
 \label{tel2}
\end{equation}
It follows that the primordial quantities in \eqref{tel1} satisfy 
\begin{equation}
n_s \simeq 0.967~, \qquad r \simeq 0.00326 \ ,
\end{equation}
which are consistent with the Planck2015 bounds \cite{Ade:2015lrj} . Putting the value of the $r$ parameter into \eqref{Vconstraint}, then implies 
\begin{equation}
V_{*}^{1/4}= 7.97 \times 10^{15}~\mathrm{GeV}  \quad \Longrightarrow \quad m = 1.55 \times 10^{13}~\mathrm{GeV} \ .
\label{tel3}
\end{equation}
 Recalling that $m$ is typical of the soft mass parameters in the $B-L$ MSSM then requires, within the context of this analysis, that
 \begin{itemize}
 
 \item {\it In order to be consistent with the Planck2015 cosmological data, supersymmetry must be broken at a high scale of ${\cal{O}}(10^{13}~\mathrm{GeV})$.}
 
 \end{itemize}

\noindent The formalism of the $B-L$ MSSM was extended to allow for an arbitrarily high SUSY breaking scale in \cite{Deen:2016vyh}. In addition, note that from \eqref{tel2}, $\psi$ must be trans-Planckian at the start of inflation. It is straightforward to show, however, that the physical fields $H_u^0$, $\nu_{3,R}$ and $\nu_{3,L}$ are all sub-Planckian during the entire inflationary epoch. 
\noindent The potential $V_{soft}(\psi)$ has already arisen in other contexts, such as supergravity models of inflaton chiral multiplets \cite{Stewart:1994ts, Kallosh:2013yoa, Ellis:2013xoa, Ellis:2014rxa}. Here, however, the inflaton is a fundamental component field in a theory of supersymmetric particle physics. Furthermore, note that

\begin{itemize}

\item {\it Our $V_{soft}$ potential arises entirely from the associated soft supersymmetry breaking quadratic term, rescaled to canonically normalize the kinetic energy.}

\end{itemize}

\begin{figure}[htbp]
   \centering
	\includegraphics[scale=0.50]{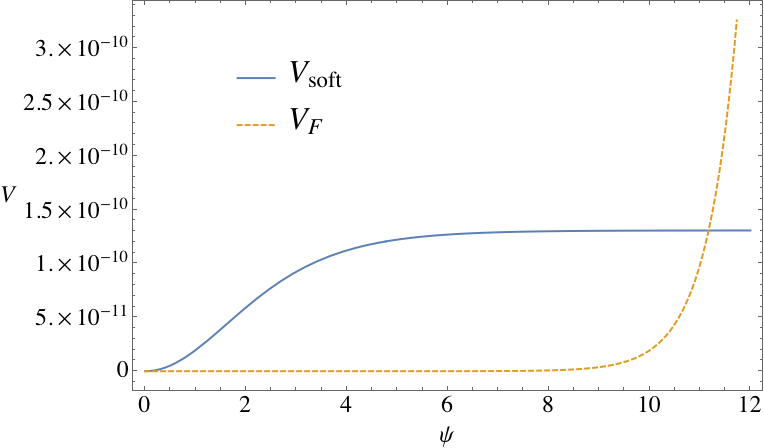}
   \caption{The blue line is a plot of $V_{soft}$ for the soft mass value $m=1.58 \times 10^{13}~\mathrm{GeV}$ in Eqn. \eqref{tel7}. The orange line is a graph of $V_{F}$ for the parameters $Y_{\nu3} \sim 10^{-12}$ and $\mu= ~1.20 \times 10^{10}~ \mathrm{GeV}$ in Eqn. \eqref{tel4}.}
   \label{potential_plot}
\end{figure}

Reintroducing the F-term potential given in the first term of \eqref{lagrange1a}, we require that it makes at most a small correction to the above results--that is, it must be  suppressed with respect to the soft mass potential\footnote{We thus avoid the ``$\eta$-problem" in supergravity models of inflation: that is, unless it is subdominant, the F-term potential would lead $\eta$ to be of $\mathcal{O}(1)$, violating the slow-roll conditions.}. In previous analyses \cite{Ovrut:2015uea}, we found that the third family sneutrino Yukawa coupling $Y_{\nu3}$ is typically very small, of order $10^{-12}$. However, to achieve sufficient suppression of $V_{F}$, the $\mu$ parameter is now forced to be at least three orders of magnitude smaller than the soft mass scale; that is,  $\mu \sim 10^{10}$ GeV. For specificity, we choose the value of $\mu$ to be close to its highest possible value:
\begin{equation}
\mu= ~1.20 \times 10^{10}~\mathrm{GeV} \ .
\label{tel4}
\end{equation}
It follows that for 60 e-foldings of inflation
\begin{equation}
\psi_{end} \simeq 1.21~, \quad \psi_* \simeq 6.25~ 
\label{tel5}
\end{equation}
and, hence, that
\begin{equation}
n_s \simeq  0.969~,\qquad r \simeq 0.00334~,
\label{tel6}
\end{equation}
again consistent with the Planck2015 data. These values correspond to the parameters
\begin{equation}
V_{*}^{1/4} = 8.07 \times 10^{15}~\mathrm{GeV}~, \quad m = 1.58 \times 10^{13}~\mathrm{GeV} \ .
\label{tel7}
\end{equation}

The potential $V_{soft}(\psi)$ is plotted as the blue line in Fig. 1 for the parameter $m$ in \eqref{tel7}.  Similarly, the F-term potential $V_{F}(\psi)$ is plotted as the dashed orange line in Fig.1 using parameter $\mu$ in \eqref{tel4}. It follows from the first term in \eqref{lagrange1a} that $V_{F}$ has a pole for $\phi_{1}=\sqrt{3}$ and, hence, using \eqref{psi}, that this function grows without bound as $\psi \rightarrow \infty$. Note that $V_{F}$ is negligible compared to $V_{soft}$ from $\psi=0$ all the way up until $\psi \sim 8$, at which point $V_{F}$ increases very  rapidly. That is, the F-term potential acts as a natural ``cut-off'' for the inflationary potential $V_{soft}$ for values of $\psi \gtrsim 8$. This gives a supersymmetric realization of the ``Inflation without Selfreproduction'' mechanism introduced in \cite{Mukhanov:2014uwa}. 

\begin{figure}[htbp]
   \centering
	\includegraphics[scale=0.50]{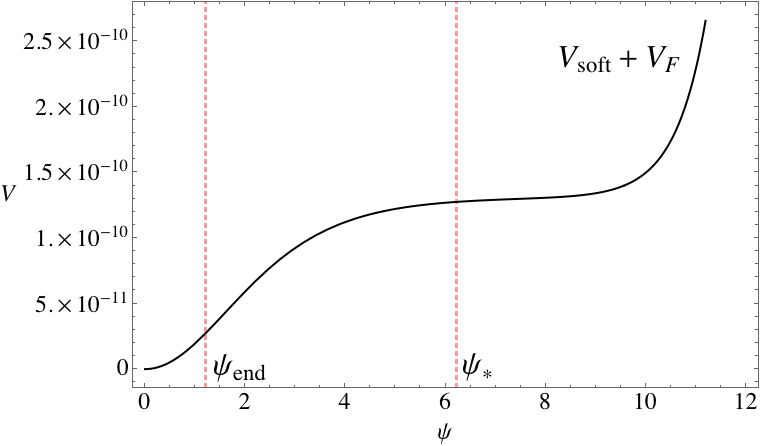}
   \caption{The black line is a graph of the potential $V_{soft}+V_{F}$ for the parameters  $m = 1.58 \times 10^{13}~\mathrm{GeV}$, $Y_{\nu3} \sim 10^{-12}$ and $\mu= ~1.20 \times 10^{10}~\mathrm{GeV}$. For these values of the parameters, the vertical red dashed lines mark $\psi_{end} \simeq 1.21$ and $\psi_{*}  \simeq 6.25$ respectively.}
   \label{plot2}
\end{figure}

The complete potential in (\ref{lagrange2}), that is, the sum of $V_{soft}+V_{F}$, is plotted in Fig. 2. It will, for suitable values of couplings $ Y_{\nu3}, \mu$ and soft mass $m$, produce a period of inflation that is consistent with current cosmological bounds obtained by Planck2015. Furthermore, these values of $m, Y_{\nu3}$ and $\mu$ can correspond to low-energy particle physics phenomenology in the $B-L$ MSSM consistent with all known low-energy data and current LHC bounds on supersymmetry, as we will demonstrate below.

\subsection{Stability}
Within the $B-L$ MSSM, our inflaton field $\phi_1$ defines a single direction in a complicated, many-dimensional field space.  For this to be a viable inflationary model, one must demonstrate that, during the inflationary epoch,  this direction is safe from displacements in field directions orthogonal to $\phi_1$. That is, no deviation away from our trajectory forces us to exit slow-roll inflation and continue down another direction in field space. 
However, we allow for displacements that lead to an orthogonal field attaining a VEV, provided this is small compared to the net field displacement of $\phi_1$ during inflation--which is of ${\cal{O}}(1)$ in Planck units. This defines our criterion for the stability of the inflationary trajectory.

In order to show that our trajectory meets this criterion, we examine the second derivative matrix of the scalar potential evaluated at each value of $\phi_1$, where the derivatives are with respect to the real and imaginary components of a given field in the $B-L$ MSSM. 
For clarity we consider only the contributions due to the F- and D-term potentials.
We first find that the second derivative matrix is block diagonal, with most of the blocks being four-by-four, involving a right-handed squark or slepton and their corresponding left-handed partner. Two exceptions arise: a six-by-six block involving the fields $\phi_2, \phi_3$ and $H_d^0$, and an eight-by-eight block involving the fields $e_{3,R}, e_{3,L}, H_u^+$ and $H_d^-$. It is clear why the larger blocks arise--any gauge invariant piece in the Lagrangian that involves the constituent fields of $\phi_1$ must involve the fields in the six-by-six and eight-by-eight blocks. 

All of the up-type squark blocks and the first and second family sneutrino blocks correspond to stable directions in field space; that is, once diagonalized, they have positive mass-squared eigenvalues. 
The down-type squarks and the first and second family selectron blocks each contain a pair of negative eigenvalues, corresponding to two unstable directions.
Diagonalizing and examining these unstable directions, we are able to conclude that, while the inflaton  may initially roll away, the D-term potential provides a sufficiently large positive contribution that the size of the resulting VEV is always of order $10^{-5}$ or less in Planck units. 
The eight-by-eight and six-by-six blocks also have positive eigenvalues. However, the fields in the six-by-six block grow a linear term for any non-zero value of $\phi_1$. This results in a set of displacements, which are again at most order $10^{-5}$.
We thus conclude that the inflationary trajectory, ignoring roll-offs that are much smaller than the distance in field space traversed during inflation, is stable.

Finally, we have redone the above computations including the quadratic and cubic soft supersymmetry breaking terms in addition to the F- and D-term potentials. The result is that the above conclusions are not changed. That is,  the inflationary trajectory, ignoring roll-offs that are much smaller--on the order of $10^{-5}$ or less--than the distance traversed by $\phi_{1}$ during inflation, is stable.

\section{Search for valid low-energy points}

In this section, we use the formalism presented in \cite{Deen:2016vyh} to statistically search the space of initial soft supersymmetry breaking parameters for those points which a) satisfy the Planck2015 data requirement in \eqref{tel7} that $m=1.58 \times 10^{13} ~ \mathrm{GeV}$ while b) simultaneously being consistent with all present low energy phenomenological data--that is, appropriate $B-L$ and EW breaking, all lower bounds on SUSY sparticles and the experimentally measured lightest neutral Higgs mass of $\sim$125 GeV. We refer the reader to \cite{Deen:2016vyh} for details of this formalism. Suffice it here to say that initial dimensional soft SUSY breaking parameters are analyzed by randomly scattering them in the interval
$[m/f,fm]$, where $f=3.3$. The results satisfying both of these requirements are shown as the ``valid'' black  points in Fig. 3.
\begin{figure}[htbp]
   \centering
	\includegraphics[scale=1.20]{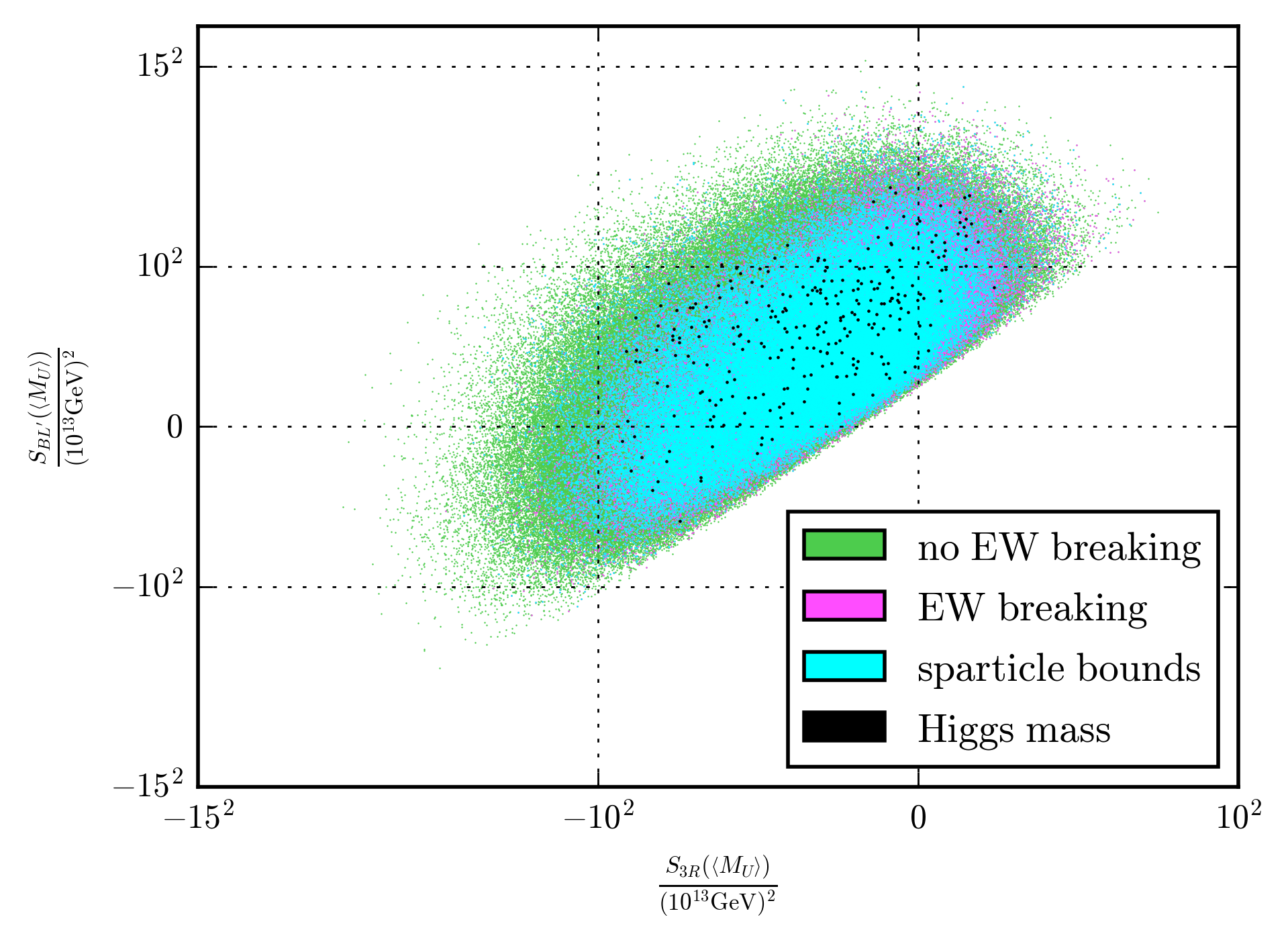}
  \caption{A scatter plot showing randomly generated points in the  $S_{3R}$-$S_{BL^\prime}$ plane. The points come from a set of 10 million randomly generated points, each constrained to be in the interval $[m/f,fm]$, where $m=1.58 \times 10^{13} ~ \mathrm{GeV}$ and $f=3.3$. The 841,952 points that break $B-L$ symmetry are shown in green. The 172,374 points that also break electroweak symmetry are shown in purple. The 109,444 points that, in addition, satisfy sparticle lower bounds are shown in cyan. Finally, the 276 points that have all the above properties and are consistent with the observed value of the lightest neutral Higgs mass are shown in black. The black points are slightly enlarged to make them more visible.}
   \label{plot3}
\end{figure}
The above results were calculated using the central experimental value of $173.21 ~ \mathrm{GeV}$ for the top quark mass \cite{Agashe:2014kda}. However, for completeness, we have redone the calculation for values of the top quark mass one, two and three sigma smaller than this value. We find, in each case, that the number of valid black points sequentially, and substantially, increases.

\section{Conclusion}

We conclude that, as stated above, there is a robust set of soft SUSY breaking initial conditions for which the $B-L$ MSSM has an inflationary epoch consistent with the Planck2015 data while simultaneously satisfying all present low energy phenomenological constraints. Inflationary scenarios are well-known to possess possible generic problems, such as the ``initial condition'' problem, the so-called ``multiverse'' problem and so on. We have made no attempt in this paper to address these generic issues--other than pointing out that our theory naturally implements the mechanism described in \cite{Mukhanov:2014uwa}. It is quite possible that, instead of inflation, the universe might have gone through a ``bounce'' from a contracting phase  to the present epoch of expansion--as discussed, for example, in \cite{Khoury:2001wf, Buchbinder:2007ad, Khoury:2010gb, Brandenberger:2016vhg}. Be that as it may, this paper demonstrates that a reasonable theory of inflation can occur in a minimal, phenomenologically acceptable $N=1$ supersymmetric theory which is softly broken at a high scale of ${\cal{O}}(10^{13}~ \mathrm{GeV})$.

We note that previous papers have attempted to use the Higgs scalar alone, both in the non-supersymmetric \cite{Bezrukov:2007ep} and supersymmetric contexts \cite{Ibanez:2014swa,Bielleman:2015lka,Einhorn:2009bh,Ferrara:2010yw} , as well as pure sneutrinos \cite{Murayama:1993xu,Ellis:2003sq,Ellis:2004hy}, as the inflaton. However, these papers have various difficulties, such as requiring non-standard coupling to gravitation and so on. None of these difficulties occur in the Sneutrino-Higgs inflation discussed in this paper.

\section*{Acknowledgments}
R. Deen and B.A. Ovrut are supported in part by DOE contract No. DE-SC0007901. A. Purves was supported in part by
DOE contract No. DE-SC0007901 when some of this work was carried out. B. Ovrut would like to thank the members of the PCTS working group ``Rethinking Cosmology'' for many helpful conversations. Finally, R. Deen is grateful to the Center for Particle Cosmology at the University of Pennsylvania for their support.

\end{document}